\def\e{{\rm e}}
\def\del{\partial}
\def\half{{1\over2}}

\def\vev#1{\langle #1 \rangle}
\def\del{\partial}
\def\half{{1\over2}}

\def\vev#1{\langle #1 \rangle}
\def\del{\partial}
\def\dslash{\del\kern-0.55em\raise 0.14ex\hbox{/}}

\def\rough#1{\raise.3ex\hbox{$#1$\kern-.75em\lower1ex\hbox{$\sim$}}}
\newcommand{\PRD}[3]{{\it Phys. Rev.} {\bf D{#1}} (19{#3}) {#2}}
\newcommand{\PRDM}[3]{{\it Phys. Rev.} {\bf D{#1}} (20{#3}) {#2}}

\newcommand{\PRLM}[3]{{\it Phys. Rev. Lett.} {\bf {#1}} {#2} (20{#3})}
\newcommand{\NPB}[3]{{\it Nucl. Phys.} {\bf B{#1}} {#2} (19{#3})}
\newcommand{\NPBM}[3]{{\it Nucl. Phys.} {\bf B{#1}} (20{#2}) {#3}}
\newcommand{\PLB}[3]{{\it Phys. Lett.} {\bf {#1}B} (19{#3}) {#2}}
\newcommand{\PLBM}[3]{{\it Phys. Lett.} {\bf B{#1}} (20{#3}) {#2}}

\newcommand{\ANN}[3]{{\it Ann. Phys. (N.Y.)} {\bf {#1}}, {#2} (19{#3})}

\newcommand{\MPL}[3]{{\it Mod. Phys. Lett.} {\bf A{#1}} (19{#3}) {#2}}
\newcommand{\MPLM}[3]{{\it Mod. Phys. Lett.} {\bf A{#1}} (20{#3}) {#2}}

\newcommand{\jhep}[3]{{\it JHEP} {\bf {#1}} (20{#2}) {#3}}

\newcommand{\tpsi}{{\tilde\psi}}
\documentclass[12pt]{article}
\usepackage[xdvi]{graphicx}
\begin{document}
\baselineskip=18pt
%%%%%%%%%%%%%%%%%%%%%%%%%%%%
\begin{titlepage}
%%%%% PREPRINT NUMBERS %%%%%%
\begin{flushright}
%28/04/2005
ROMA1/1425/06\\
OU-HET-556/2006
%hep-th/0305xxx
\end{flushright}
\vspace{1cm}
%%%%%%%%%%%%%%%%%%% TITLE %%%%%%%%%%%%%%%%%%
\begin{center}{\Large\bf 
Effects of Bulk Mass in Gauge-Higgs Unification}
\end{center}
%%%%%%%%%%%%%%%% AUTHORS %%%%%%%%%%%%%%%%%%%%%%%
\vspace{1cm}
\begin{center}
Nobuhito Maru$^{(a)}$
\footnote{E-mail: Nobuhito.Maru@roma1.infn.it (N.Maru)} and
Kazunori Takenaga$^{(b)}$
\footnote{E-mail: takenaga@tuhep.phys.tohoku.ac.jp (K.Takenaga)\\
Present address : Department of Physics, Tohoku University, Sendai,
980-8578 Japan.}
\end{center}
%%%%%%%%%%%%%%%%%%%%%%% AFFILIATION %%%%%%%%%%%%
\vspace{0.2cm}
\begin{center}
%\small
${}^{(a)}$ {\it Dipartimento di Fisica, Universit\`a di Roma
``La Sapienza'' \\ 
and INFN, Sezione di Roma, P.le Aldo Moro, I-00185 Roma,Italy}
\\[0.2cm]
${}^{(b)}$ {\it Department of Physics, Osaka University, 
Toyonaka, Osaka 560-0043, Japan}
%%%%%
%%%%%%%
\end{center}
%%%%%%%%%%%%%%%%%% ABSTRACT %%%%%%%%%%%%%%%
\vspace{1cm}
\begin{abstract}
We study effects of bulk mass on electroweak symmetry 
breaking and Higgs mass in the scenario of five 
dimensional $SU(3)$ gauge-Higgs unification defined 
on $M^4\times S^1/Z_2$. The asymptotic form of effective
potential for the Higgs field is obtained, from which 
a transparent and useful expression for the Higgs mass is 
found. The small vacuum expectation values (VEV) for 
Higgs field can be realized by choosing 
bulk mass parameters approriately for 
a fixed set of matter content. The 
bulk mass for periodic fermion 
field, in general, has effects to make the Higgs mass 
less heavy. On the other hand, the bulk mass for antiperiodic 
field does not directly affect the Higgs mass, but it 
contributes to 
%lift 
increase or decrease the Higgs mass, depending
on how small the VEV is induced due to the antiperiodic 
field. We give numerical examples to confirm these effects, 
in which the role of the bulk mass is also 
%manifestly 
definitely clear.
\end{abstract}
\end{titlepage}
%%%%%%%%%%
%\tableofcontents
%%%%%%%%%%%%
\newpage
%%%%%%%%%%%%%%%% INTRODUCTION %%%%%%%%%%%%%%%
\section{Introduction}
The idea of gauge-Higgs unification, originally proposed by 
Manton \cite{manton} and Fairlie \cite{fair}, is one of the 
attractive ideas for resolving the (little) gauge hierarchy 
problem \cite{gaugehiggs1}. In the gauge-Higgs 
unification, the Higgs field is unified as extra 
components in higher dimensional
gauge field and is related to the Wilson line phases 
along the compactified space direction. 
The electroweak symmetry is broken through the dynamics 
of the Wilson line phases (Hosotani 
mechanism \cite{hosotani}), and the Higgs field
becomes massive, though it is massless at the tree-level, 
which is obtained by the effective potential 
induced by radiative corrections. The gauge-Higgs unification,
including the dynamics of the Wilson line phases, has been
studied from various points of 
view in the past \cite{gaugehiggs2, models, hmoy}.
\par
%%%%%%%%%%%%
The potential for the Higgs field generated at quantum 
level has a non local form, reflecting the 
nonlocal nature of the Wilson line 
phases \cite{masiero}, so that it is free from 
ultraviolet effects. That is why the mechanism can provide 
an alternative solution to the gauge hierarchy problem. 
Moreover, the Higgs sector in the gauge-Higgs unification has no 
free parameters. Once we fix matter content, the Higgs mass is
calculable. The size of the Higgs mass, generally 
speaking, is small compared with the gauge boson mass 
because it is generated by the radiative correction, which 
is similar to the Coleman-Weinberg mechanism \cite{cw}. 
In order for the scenario to be valid, one needs the heavy 
Higgs mass consistent with the experimental lower bound.
\par
%%%%%%%%%%%
For the heavy Higgs mass, it has been known 
that there are two points that 
have to be satisfied. The first point 
is that the vacuum expectation values (VEV) for the 
Higgs field must be small enough, and the second point is 
that one needs the matter belonging to the higher 
dimensional representation under the gauge 
group \footnote{It has been also reported that effects of 
the breakdown of the Lorentz invariance also enhance the 
Higgs mass \cite{roma2}.} \cite{hty2, csaki}. The first 
point is realized by choosing an appropriate matter 
content in such a way that the
coefficient of the negative mass term of the 
induced Higgs potential is very small \cite{hty2}. 
\par
%%%%%%% 
Bulk mass plays an important role when we discuss 
the gauge symmetry breaking patterns through the dynamics of the
Wilson line phases \cite{takenaga, hty1} and fermion mass 
spectrum \cite{roma1, csaki} in the gauge-Higgs unification. 
The bulk mass is a necessary 
ingredient for model building based on the gauge-Higgs 
unification even though, as a result, it introduces free 
parameters in the Higgs sector. 
%Neverthless, 
Therefore, it is important 
to understand the role and effect of the bulk mass in the 
gauge-Higgs unification \footnote{The role of the bulk mass is 
also studied from a point of view of low-energy effective 
theory of the gauge-Higgs unification \cite{hmoy}}. 
\par
%%%%%%%%%%
In this paper we study the effect of the bulk 
mass in nonsupersymmetric five 
dimensional $SU(3)$ gauge-Higgs 
unification theory, where an extra spatial coordinate is
compactified on an orbifold $S^1/Z_2$. As is well 
known, the fermion bulk mass term in five dimensions
is parity odd under the reflection of the coordinate of the
extra dimension. In order to have the parity even 
bulk mass term, we introduce a pair of the fields, each 
satisfying periodic or antiperiodic boundary
condition in this paper. The effect of the periodic 
field in the gauge-Higgs unification is different from 
that of the antiperiodic field.
%%%%
%We point out that the electroweak symmetry breaking and the Higgs
%mass are sensitive for the bulk mass parameter. 
%
%%%%%%%. 
We give an aymptotic form of the effective potential, from which
the effect is definitely clear and, also from which a transparent 
and useful expression for the Higgs mass is found. 
\par
%%%%%%%
It is possible to have the small VEV for the Higgs field by 
choosing the bulk mass parameters appropriately for a 
fixed matter content. Namely, the bulk mass for the 
antiperiodic field is important 
to control the magnitude of the VEV. 
Since the effective potential depends on the bulk mass parameters, 
%accordingly, 
the Higgs mass also depends on them. 
The size of the Higgs mass is mainly controlled 
by the logarithmic factor arising only from the periodic fields. 
The bulk mass for the periodic field, in general, has effects 
to make the Higgs mass less heavy. 
In order to enhance the Higgs mass, one needs smaller 
VEV for moderate size of the bulk mass. 
On the other hand, the bulk mass itself for the antiperiodic field 
has tiny effects on the Higgs mass, but the field has a crucial role 
to make the VEV large (small), which 
accordingly can increases (decreases) the Higgs mass.
%%%
%does not have 
%But the bulk mass for the antiperiodic field does not
%directly affect the Higgs mass unlike the one for periodic
%field, on which the Higgs mass explicitly depends.
%The bulk mass is, in general, suppresses the size of the Higgs 
%mass though it is possible to realize the small VEV for 
%the Higgs field. An interesting possibility is that this is 
%not always the case. 
%%%%%%
We will present numerical examples, in which the effect of the 
bulk mass is definitely clear.
\par
%%%%%%%%%%%%%
In the next section we present our model and study the electroweak
symmetry breaking in the gauge-Higgs unification, paying attention to 
the size of the VEV for the Higgs field and the Higgs mass. 
In section 3, some numerical examples are given. 
Section 4 is devoted to conclusions and discussions. 
The derivation of asymptotic form of the potential is breifly 
summarized in an appendix. 
%%%%%%%%%%%%%%
\section{Bulk mass parameter and effective potential}
We consider nonsupersymmetric gauge 
theory on $M^4\times S^1/Z_2$ with the gauge
group $SU(3)$, where $M^4$ is the four dimensional Minkowski
space-time and $S^1/Z_2$ is an orbifold \footnote{The gauge-Higgs
unification in supersymmetric
gauge theory with bulk mass is also studied in \cite{hty2}.}. One
needs to specify
boundary conditions of fields for the $S^1$ direction and the 
two fixed points located at $y=0, \pi R$. These boundary 
conditions are defined by the two 
matrices, $P_0=P_1=\mbox{diag}.(-1,-1,1)$, from which the original 
gauge symmetry is broken down 
to $SU(2)\times U(1)$ \cite{orbi}. And the Higgs 
field is embedded in the zero 
modes of the extra component of the gauge field $A_y$ as
\begin{equation}
A_y^{(0)}=\half 
\pmatrix{ & &A_y^4-iA_y^5  \cr 
          & &A_y^6-iA_y^7  \cr
         {\rm c.c.}&{\rm c.c.}& \cr}
\quad\mbox{where}\quad
\Phi\equiv \sqrt{2\pi R}\pmatrix{A_y^4 -i A_y^5 \cr A_y^6 -i A_y^7 \cr}, 
\label{shiki1}
\end{equation}
where $R$ is a radius of the $S^1$. $\Phi$ transforms as a
doublet under the $SU(2)$. 
\par
%%%%%%%%%
The VEV of the Higgs field 
is parametrized by
\begin{equation}
\vev{A_y^{(0)}}={a\over {g_4 R}}{\lambda^6\over 2}
\equiv A_y^{6(0)}{\lambda^6\over 2}.
\label{shiki2}
\end{equation}
Here, $g_4$ stands for the four dimensional gauge coupling defined by
the original five dimensional coupling $g_5$ by $g_4\equiv
g_5/\sqrt{2\pi R}$, and $a$ is a real constant. 
The VEV $a$ is related to the Wilson line phase and
determines the patterns of the gauge symmetry breaking,
\begin{eqnarray}
W &=& {\cal P} \mbox{exp} \left(ig \oint_{S^1}dy \vev{A_y} \right)
= \pmatrix{
1 & 0 & 0 \cr 
0 & \cos(\pi a_0)  & i\sin(\pi a_0) \cr
0 & i\sin(\pi a_0) & \cos(\pi a_0) }\qquad (a_0~~\mbox{mod}~2)
\nonumber\\
&=& \left\{
\begin{array}{lll}
SU(2)\times U(1) & \mbox{for} & a_0=0, \\[0.1cm]
U(1)^{\prime}\times U(1)& \mbox{for} & a_0=1, \\[0.1cm]
U(1)_{em} & \mbox{for} & \mbox{otherwise}.
\end{array}\right.
\label{shiki3}
\end{eqnarray}
$a_0$ represents the value of $a$ at the minimum of the potential. 
In order to determine the values of the Wilson line 
phase $a$, one usually evaluates the effective potential 
for the phase and minimizes it. Then, the Higgs mass is calculated 
by the second derivative of the effective potential at the minimum. 
Since $a_0$ is related to weak gauge boson masses 
whose size is given by $a_0/R$, if one requires $1/R \sim $ a few TeV, 
then, $a_0$ must be of order of $O(10^{-2})$. 
This is a strong constraint that should be satisfied 
in gauge-Higgs unification.
\par
%%%%%%%%%%%%%
We consider the matter field with bulk mass term. The 
bulk mass term for fermion in five dimensions is odd 
under the parity transformation, $y\rightarrow -y 
(\pi R-y \rightarrow \pi R+y)$. We need parity even
mass term for the consistency of the $Z_2$ orbifolding. The parity even
mass term is introduced by the coordinate dependent mass term 
such as, for example, $M(-y)=-M(y)$, where 
$M(y)=\epsilon(y)M(\epsilon(y)$ is the step function). 
%
%In this
%case, the mass eigenvalues do not have analytic form \footnote{This
%point is extensively studied in the forthcoming paper \cite{mt}.}. 
%
The other way for obtaining the parity even mass term is to introduce a
pair of the fields, $\psi_+$ and $\psi_-$ whose parity is different
to each other, $\psi_{\pm}(-y)=\pm \psi_{\pm}(y)$. Then, a parity 
even mass term is constructed like $M{\bar\psi}_{+}\psi_{-}$. 
In this paper we follow this case. 
\par
%%%%%%%%%%%%%%
Let us suppose that $\psi^{(\pm)}$ and ${\tilde\psi}^{(\pm)}$ belong to
the fundamental representation under the $SU(3)$ gauge group and satisfy
the following boundary conditions \footnote{Notation used in this 
section is the same as those in \cite{hty2}.},
%%%%%%%%%%%
\begin{eqnarray}
\mbox{type~I}&;&\left\{
\begin{array}{l}
\psi^{(+)}(-y)=P_0~i\Gamma^y~\psi^{(+)}(y)\\[0.3cm]
\psi^{(-)}(-y)=-P_0~i\Gamma^y~\psi^{(-)}(y),
\end{array}\right.
\left\{
\begin{array}{l}
\psi^{(+)}(\pi R-y)=P_1~i\Gamma^y~\psi^{(+)}(\pi R+y)\\[0.3cm]
\psi^{(-)}(\pi R-y)=-P_1~i\Gamma^y~\psi^{(-)}(\pi R+y).
\end{array}\right.\label{shiki4}
\\[0.3cm]
%\end{equation}
%%%%%%%%%%%%%%
\mbox{type~II}&;&\left\{
\begin{array}{l}
\tpsi^{(+)}(-y)=P_0~i\Gamma^y~\tpsi^{(+)}(y)\\[0.3cm]
{\tpsi}^{(-)}(-y)=-P_0~i\Gamma^y~{\tpsi}^{(-)}(y),
\end{array}\right.
\left\{
\begin{array}{l}
\tpsi^{(+)}(\pi R-y)=-P_1~i\Gamma^y~\tpsi^{(+)}(\pi R+y)\\[0.3cm]
{\tpsi}^{(-)}(\pi R-y)=P_1~i\Gamma^y~{\tpsi}^{(-)}(\pi R+y).
\end{array}\right.
\label{shiki5}
\end{eqnarray}
We can also consider the boundary conditions where the overall sign 
is different from the above boundary conditions, but these 
contributions to the effective 
potential are the same as
those defined in Eqs. (\ref{shiki4}) and (\ref{shiki5}). It is easy to
see that a pair $(\psi^{(+)}, \psi^{(-)}) ((\tpsi^{(+)},
\tpsi^{(-)})$ can have the parity
even and gauge invariant mass term,
$M{\bar\psi}^{(\pm)}\psi^{(\mp)} (M{\bar\tpsi}^{(\pm)}\tpsi^{(\mp)})$.
%%%%%%%%%
%The Lagrangian for the
%fields is given by
%\begin{eqnarray}
%{\cal L}&=&
%{\bar\psi}^{(+)}i\Gamma^{\hmu}D_{\hmu}\psi^{(+)}
%+{\bar\psi}^{(-)}i\Gamma^{\hmu}D_{\hmu}\psi^{(-)}
%-M{\bar\psi}^{(+)}\psi^{(-)}+h.c.\nonumber\\
%&+&{\bar\tpsi}^{(+)}i\Gamma^{\hmu}D_{\hmu}{\tpsi}^{(+)}
%+{\bar\tpsi}^{(-)}i\Gamma^{\hmu}D_{\hmu}{\tpsi}^{(-)}
%-M{\bar\tpsi}^{(+)}{\tpsi}^{(-)}+h.c.
%\label{shiki6}
%\end{eqnarray} 
%%%%%%%
Taking into account the boundary conditions (\ref{shiki4}) 
and (\ref{shiki5}), $\psi^{(\pm)}$ is expanded in terms 
of $\cos({n\over R}y)$ and $\sin({n \over R}y)$, while ${\tpsi}^{(\pm)}$
is expanded in terms of $\cos({{(n+\half)y}\over R})$ 
and $\sin({{(n+\half)y}\over R})$. We call the field 
with the expansion $\cos({n\over R}y)(\sin({n\over R}y))$ periodic
field, while that with the
expansion $\cos({{(n+\half)y}\over R})(\sin({{(n+\half)y}\over
R}))$ antiperiodic field.
\par
%%%%%%%%%%
The contributions to the effective
potential from the fermions $\psi^{(+)}$ and $\psi^{(-)}$ is given by
\begin{equation}
V_{eff}^{I}=-2^{[{5\over 2}]}N_I^{pair}~(1+1)~{1\over L}~\half
\sum_{n=-\infty}^{\infty}\int{{d^4p_E}\over{(2\pi)^4}}~
{\rm ln}\left[p_E^2 + \left({{n+{a\over 2}}\over R}\right)^2 + M^2 \right],
\label{shiki7}
\end{equation}
where $N_I^{pair}$ stands for the number of the pair $(\psi^{(+)},
\psi^{(-)})$. The overall minus sign comes 
for the Fermi statistics. 
$p_E$ denotes Euclidean momentum. 
Likewise, we obtain from the
pair $(\tpsi^{(+)},~\tpsi^{(-)})$ that
\begin{equation}
V_{eff}^{II}=-2^{[{5\over 2}]}N_{II}^{pair}~(1+1)~{1\over L}~\half
\sum_{n=-\infty}^{\infty}\int{{d^4p_E}\over{(2\pi)^4}}~
{\rm ln}
\left[p_E^2 + \left({{n+{a\over 2} -\half}\over R}\right)^2 
+ M^2 \right],
\label{shiki8}
\end{equation}
where  $N_I^{pair}$ stands for the number of 
the pair $(\tpsi^{(+)}, \tpsi^{(-)})$.
According to the usual prescription \cite{pomarol}, it is easy to 
evaluate Eqs. (\ref{shiki7}) and (\ref{shiki8}) as
%to give
\begin{eqnarray}
&&\half{1\over L}\sum_{n=-\infty}^{\infty}
\int{{d^4p_E}\over{(2\pi)^4}}~
{\rm ln}\left[p_E^2 + \left({{n+Qa -{\delta\over 2}}\over
R}\right)^2 
+ M^2 \right]
\nonumber\\
&=&-{2\over{(2\pi)^{5\over 2}}}
\sum_{n=1}^{\infty}\left(M\over {nL}\right)^{5\over 2}
K_{5\over 2}(nLM)\cos\left[2\pi n\left(Qa-{\delta\over
2}\right)\right]
\nonumber\\
&=&
-{3\over{4\pi^2}}{1\over L^5}\sum_{n=1}^{\infty}
{1\over n^5}\left(1+nz+{{n^2z^2}\over 3}\right)\e^{-nz}
\cos\left[2\pi n\left(Qa-{\delta\over 2}\right)\right],
\label{shiki9}
\end{eqnarray}
where we have ignored the irrelevant constant. 
$\delta$ takes $0(1)$ for (anti)periodic field.
%depending on the periodicity of the field.
We have defined the dimensionless variable $z\equiv ML$ and also 
used the fact that the modified Bessel function $K_{\nu}(x)$ can be
rewritten as
\begin{equation} 
K_{5\over 2}(x)=
3\left({\pi\over {2x^5}}\right)^{\half}
\left(1+x+{x^2\over 3}\right)\e^{-x}.
\label{shiki10}
\end{equation}
Then, the effective potential from the matter is summarized into the
form as
\begin{equation}
V_{eff}(a, z, \delta)=(-1)^{F+1}N_{deg}(2N_{pair}){3\over{4\pi^2L^5}}
f(Qa, z, \delta),
\label{shiki11}
\end{equation}
where
\begin{equation}
f(Qa, z,\delta)=\sum_{n=1}^{\infty}{1\over n^5}
\left(1+nz +{{n^2z^2}\over 3}\right)\e^{-nz}
\cos\biggl[2\pi n\Bigl(Qa-{\delta\over 2}\Bigr)\biggr].
\label{shiki12}
\end{equation}
$N_{deg}$ is the on-shell degrees of freedom of 
the concerned field, and $F$ stands for 
the fermion number. Let us note that the factor $2$ in
$2N_{pair}$ comes from $\psi^{(+)}$ and $\psi^{(-)}$ in the 
pair $(\psi^{(+)}, \psi^{(-)})$. 
\par
%%%%%%%%%%%%%
The ``charge'' $Q$ is just the magnitude of the $SU(2)$
spin $j$. Any representation of the $SU(3)$ is 
decomposed in terms of the irreducible representation of $SU(2)$. 
The fundamental representation of $SU(3)$, for example, is decomposed 
by an $SU(2)$ doublet and a singlet, so that $Q=j=1/2$ from the doublet
contribution, and the singlet contribution does not depend on 
the order parameter $a$. 
%Let us also note $\delta$ 
%takes $0, 1$, depending on the periodicity of the field.
\par
%%%%%%%%%%
%Taking into account 
Following the previous work done by the authors \cite{mt2}, 
we introduce the matter fields whose flavor numbers are specified by
\begin{equation}
(N_{adj}^I,N_{fd}^I, N_{adj}^{(+)s}, N_{fd}^{(+)s};
N_{adj}^{II}, N_{fd}^{II}, N_{adj}^{(-)s}, N_{fd}^{(-)s}).
\label{shiki13}
\end{equation}
Here, we have also introduced the scalar field with 
the $\eta=-1$ parity \footnote{Definition for the $\eta$ parity was
given in \cite{haba}}, which is essentially the same as the
antiperiodic field. The mode expansion of the scalar field with
$\eta=+1 (-1)$ is the same as those of the fermions with type I (II)
boundary conditions. Then, the effective 
potential for these flavor number of fields (\ref{shiki13}) is given by
\begin{eqnarray}
{\bar V}_{eff}(a)={V_{eff}\over{3/4\pi^2 L^5}}&=&
-3\biggl(f(2a, 0, 0) +2f(a,0, 0)\biggr)\nonumber\\
&+&4(2N_{adj}^I)\biggl(f(2a, z_{adj}^{(+)}, 0) + 
2f(a, z_{adj}^{(+)}, 0)\biggr) 
\nonumber\\
&+&4(2N_{adj}^{II})\biggl(f(2a, z_{adj}^{(-)}, 1) 
+ 2f(a, z_{adj}^{(-)}, 1)\biggr) 
\nonumber\\
&+&4(2N_{fd}^I) f(a, z_{fd}^{(+)}, 0) 
+4(2N_{fd}^{II})f(a, z_{fd}^{(-)}, 1) \nonumber\\
%%%%%%%%
&-&dN_{adj}^{(+)s}\biggl(f(2a, z_{adj}^{(+)s}, 0) 
+ 2f(a, z_{adj}^{(+)s}, 0)\biggr) 
\nonumber\\
&-&dN_{adj}^{(-)s}\biggl(f(2a, z_{adj}^{(-)s}, 1) 
+ 2f(a, z_{adj}^{(-)s}, 1)\biggr) 
\nonumber\\
&-&(2N_{fd}^{(+)s}) f(a, z_{fd}^{(+)s}, 0) 
-(2N_{fd}^{(-)s})f(a, z_{fd}^{(-)s}, 1).
\label{shiki14}
\end{eqnarray}
The first line in Eq.(\ref{shiki14}) comes from the gauge sector. 
The factor $d$ in the adjoint scalar contributions takes $1(2)$ 
%or $2$, depending on 
for the real (complex) field. 
\par
%%%%%%%%%%%%%%%%%
In order to see the role of the bulk mass clearly, let us study the 
limit $z \ll 1$ and obtain the asymptotic form of $f(Qa, z, \delta)$. 
Taking also into account the smallness of the VEV 
for the Higgs field $a$, one can utilize the formulae 
given in the appendix. Then, we find that
\begin{eqnarray} 
f(x, z, 0)&\simeq &\zeta(5) -{\zeta(3)\over 6}z^2 +{z^4\over {32}}
-{\zeta(3)\over 2}x^2+{1\over{48}}(7x^2z^2 + {{25}\over 6}x^4) \nonumber\\
&-&{1\over{48}}(x^2 + z^2)^2{\rm ln}(x^2+z^2),
\label{shiki15}
\\
f(x, z, 1)&\simeq &-{{15}\over {16}}\zeta(5) +{\zeta(3)\over 8}z^2 +
{3\over 8}\zeta(3)x^2-{{{\ln}2}\over{24}} (x^2+z^2)^2,
\label{shiki16}
\end{eqnarray}
where $x\equiv 2\pi Qa$, and we have ignored higher order terms. 
The equation (\ref{shiki15}) is the same 
as that obtained in \cite{hmoy}.
\par
%%%%%%%%%%%%%
Let us apply the formulae (\ref{shiki15}) and (\ref{shiki16}) to the 
effective potential (\ref{shiki14}). We 
obtain, apart from irrelevant constant, that
\begin{eqnarray}
{\bar V}_{eff}&\simeq &
-{{\pi^2 B}\over 2}a^2+{{\pi^4 C}\over{48}}a^4
+{9\pi^4\over 8}a^4{\rm ln}(\pi a)^2
\nonumber\\
&-&{1\over{48}}
\biggl(
4(2N_{adj}^{I})
\biggl[L(2a,z_{adj}^{(+)})+ 2L(a,z_{adj}^{(+)})\biggr]
+4(2N_{fd}^{I})L(a, z_{fd}^{(+)})\nonumber\\
&-&dN_{adj}^{(+)s}
\biggl[L(2a,z_{adj}^{(+)s})+ 2L(a,z_{adj}^{(+)s})\biggr]
-2N_{fd}^{(+)s}L(a, z_{fd}^{(+)})
\biggr),
\label{newshiki1}
\end{eqnarray}
where 
\begin{equation}
L(a,z)\equiv \left((\pi a)^2+z^2\right)^2{\rm ln}
\left[(\pi a)^2+z^2\right].
\label{newshiki2}
\end{equation}
The coefficients $B, C$ are given by
\begin{eqnarray}
B&\equiv &
\zeta(3)\Biggl[48N_{adj}^{I}+8N_{fd}^{I}
+{3\over 4}\biggl(6dN_{adj}^{(-)s}+2N_{fd}^{(-)s}\biggr)
\nonumber\\
&-&{3\over 4}\biggl(48N_{adj}^{II}+8N_{fd}^{II}\biggr)
-\biggl(6dN_{adj}^{(+)s}+2N_{fd}^{(+)s}+18\biggr)\Biggr]
\nonumber\\
&-&{7\over{24}}
\biggl(48N_{adj}^{I}z_{adj}^{(+)2}+8N_{fd}^{I}z_{fd}^{(+)2}\biggr)
+{{\rm ln}2\over 6}
\biggl(48N_{adj}^{II}z_{adj}^{(-)2}+8N_{fd}^{II}z_{fd}^{(-)2}\biggr)
\nonumber\\
&+&{7\over{24}}
\biggl(6dN_{adj}^{(+)s}z_{adj}^{(+)s2}+2N_{fd}^{(+)s}z_{fd}^{(+)s2}\biggr)
-{{\rm ln}2\over 6}
\biggl(6dN_{adj}^{(-)s}z_{adj}^{(-)s2}+2N_{fd}^{(-)s}z_{fd}^{(-)s2}\biggr),
%%%%
%&+&48N_{adj}^{I}\biggl(\zeta(3)-{7\over 24}z_{adj}^{(+)2}\biggr)
%+8N_{fd}^{I}\biggl(\zeta(3)-{7\over 24}z_{fd}^{(+)2}\biggr)
%\nonumber\\
%&+&6dN_{adj}^{(-)s}\biggl({3\over 4}\zeta(3)-
%{{\rm ln}2\over 6}z_{adj}^{(-)s2}\biggr)
%+2N_{fd}^{(-)s}\biggl({3\over 4}\zeta(3)-
%{{\rm ln}2\over 6}z_{fd}^{(-)s2}\biggr)
%\nonumber\\
%&-&48N_{adj}^{II}\biggl({3\over 4}\zeta(3)-
%{{\rm ln}2\over 6}z_{adj}^{(-)2}\biggr)
%-8N_{fd}^{II}\biggl({3\over 4}\zeta(3)-
%{{\rm ln}2\over 6}z_{fd}^{(-)2}\biggr)
%\nonumber\\
%&-&6dN_{adj}^{(+)s}\biggl(\zeta(3)-{7\over 24}z_{adj}^{(+)s2}\biggr)
%-2N_{fd}^{(+)s}\biggl(\zeta(3)-{7\over 24}z_{adj}^{(+)s2}\biggr)
%-18\zeta(3),
%%%%%%%
\label{newshiki3}
\\
C&\equiv & {25\over 6}
\biggl(72(2N_{adj}^{I})
+4(2N_{fd}^{I})-18dN_{adj}^{(+)s}-2N_{fd}^{(+)s}-54
\biggr)
\nonumber\\
&+&
2{{\rm ln}2} 
\biggl(
18dN_{adj}^{(-)s}+2N_{fd}^{(-)s}+54
-72(2N_{adj}^{II})-4(2N_{fd}^{II})
\biggr).
%%%%%
%+dN_{adj}^{(-)s}{{\rm ln}2\over 24}18
%+2N_{fd}^{(-)s}{{\rm ln}2\over 24}
%\nonumber\\
%&-&4(2N_{adj}^{II}){{\rm ln}2\over 24}18 
%-4(2N_{fd}^{II}){{\rm ln}2\over 24}
%-dN_{adj}^{(+)s}{25\over 6}
%-2N_{fd}^{(+)s}{25\over 6}\nonumber\\
%&+&48 (2{\rm ln}2)-54{25\over 6}
%%%%%%%
\label{newshiki4}
\end{eqnarray}
\par
%%%%%%%%%
If we do not introduce the bulk mass at all, in order to 
obtain the small VEV for the Higgs field, we need to 
choose the matter content in such a way that the 
coefficient $B$ is almost cancelled \cite{hty2}, that is,
the cancellation between the first 
and the second lines in Eq.(\ref{newshiki3}) is required. 
Here, thanks to the bulk mass
parameters, we can make the coefficient $B$ small 
by choosing the parameters appropriately  
even for a fixed matter content. 
%%
%Namely, if we look at the fermion fields
%satisfying the type I and II boundary conditions, we see that 
%the bulk mass of the field with $\delta=1$, that is, the 
%type II boundary condition makes the VEV for the Higgs field small.
%%
\par
%%%%%%%%%
%We note that the bulk mass of the field 
%with $\delta=1$ tends to break the gauge symmetry 
%due to the negative contributions for the 
%coefficient $B$, which is interesting because the contribution
%from the massless bulk field 
%with $\delta=1$ to the coefficient is quite oppisite. 
In general, we see that the massive bulk 
fermion with $\delta=0$ has effects to restore the gauge 
symmetry because of the positive mass 
term $7z^2x^2/48$ in Eq.(\ref{shiki15}), while 
from Eq. (\ref{shiki16}), the massive bulk fermion 
with $\delta=1$ has effects to break the gauge symmetry 
due to the negative mass term $({\rm ln}2/12) z^2x^2$. 
Let us also note that the boson with $\eta=+(-)$ parity 
has the same role as the fermion with $\delta=1(0)$. 
\par
%%%%%%%%%
The Higgs mass is obtained from the second derivative 
of the effective potential at the minimum. The asymptotic
expression for the Higgs mass
is given, using (\ref{newshiki1})
and the relation $a_0/g_4R=v$ \cite{ghrelation}, by
\begin{equation}
m_H^2=g_4^4 {3\over{64\pi^6}}\left({v\over a_0}\right)^2
{{\del^2{\bar V}_{eff}}\over{\del a^2}}\bigg|_{a=a_0},
\label{newshiki5}
\end{equation}
where
\begin{eqnarray}
{{\del^2{\bar V}_{eff}}\over{\del a^2}}
&=&{8 \pi^4 a_0^2}
\biggl[
{{-4(2N_{adj}^{I})}\over {48}}
\biggl(2^4 H(2a, z_{adj}^{(+)}) + 2\times 1^4 H(a, z_{adj}^{(+)})\biggr)
\nonumber\\
&-&{{4(2N_{fd}^{I})}\over {48}} H(a, z_{fd}^{(+)})
+{9\over 8}\biggl({\rm ln}(\pi^2 a^2)+{3\over 2}\biggr)
+{C\over 48}
\nonumber\\
&+&{{dN_{adj}^{(+)s}}\over {48}}
\biggl(2^4 H(2a, z_{adj}^{(+)s}) +2\times 1^4 H(a,z_{adj}^{(+)s})\biggr)
\nonumber\\
&+&{{2N_{fd}^{(+)s}}\over {48}} 1^4  H(a, z_{fd}^{(+)s})
\biggr],
\label{shiki17}
\end{eqnarray}
where
\begin{equation}
H(a, z)={\rm ln}\left[(\pi a)^2+z^2\right]+{3\over 2}
\label{newshiki6}
\end{equation}
\par
%%%%%%%
We observe that the Higgs mass 
depends on the magnitude of the $SU(2)$ spin like 
$(2Q)^4 (=2^4, 1^4)$, in front of $H(a,z)$, so that
the matter with the higher dimensional representation 
under the gauge group enhances the size of the Higgs mass 
\cite{csaki, hty2}. 
\par
%%%%%%%%%
A remarkable feature is that the bulk mass itself of the 
field with $\delta=1$ does not affect the Higgs mass 
directly, 
%so that one may think it does not affect the 
%Higgs mass at all. This is not true. The 
%field with $\delta=1$ 
but it contributes to increase or decrease the Higgs mass 
only through the magnitude of the VEV for the Higgs field. 
%
%because 
%
The bulk mass for the field with $\delta=1$ 
has a role of changing the magnitude. We will 
explicitly show such examples in the next section. 
%%%%%
%Let us note again that the logarithmic dependence 
%comes form only the fermions with type I boundary condition 
%and the scalars with $\eta=+$ parity and dominantly controls the size
%of the Higgs mass. 
%%%
%%%%%%%%%%
\par
On the other hand, the bulk mass for the field with $\delta=0$  
can have sizable effect on the Higgs mass. First of all, 
let us note that 
the large logarithmic factor due to the small VEV $a_0 \ll 1$ 
comes from only the field with $\delta=0$, 
as seen from Eq.(\ref{shiki17}). 
In our numerical analyses given below, 
we do not consider the scalar fields with $\delta=0$, 
so that the Higgs mass is dominantly controlled by the fermion 
fields with $\delta=0$ (and the gauge field). 
%The large logarithm factor is important to obtain the heavy Higgs mass. 
However, the bulk mass for the field with $\delta=0$ 
modifies the argument of the logarithm, 
which implies that the bulk mass tends to decrease 
the size of the Higgs mass in general. 
Therefore, in order to avoid the light Higgs mass, 
one needs to have smaller values of the 
VEV $a_0$ for moderate size of the bulk mass, 
which overcomes the $z^2$-suppression in the 
argument of the logarithm.
\par
%%%%%%%%%%%%%%%%%%%%%%%%%%%
\section{Numerical results}
%%%%%%%%%%%%%%%%%%%%%%%%%%%
In this section, 
let us show numerical results for the following three cases,
\begin{eqnarray}
(\mbox{A})~~(N_{adj}^I,N_{fd}^I, N_{adj}^{(+)s}, N_{fd}^{(+)s}; N_{adj}^{II}, 
N_{fd}^{II}, N_{adj}^{(-)s}, N_{fd}^{(-)s})&=&(1,1,0,0;1,1,1,0),
\\
(\mbox{B})~~(N_{adj}^I,N_{fd}^I, N_{adj}^{(+)s}, N_{fd}^{(+)s}; N_{adj}^{II}, 
N_{fd}^{II}, N_{adj}^{(-)s}, N_{fd}^{(-)s})&=&(1,1,0,2;1,1,2,0),
\\
(\mbox{C})~~(N_{adj}^I,N_{fd}^I, N_{adj}^{(+)s}, N_{fd}^{(+)s}; N_{adj}^{II}, 
N_{fd}^{II}, N_{adj}^{(-)s}, N_{fd}^{(-)s})&=&(1,1,0,0;1,1,0,3).
\label{shiki18}
\end{eqnarray}
\par
%%%%%%%%% 
We first present the numerical results for the case (A) in the table 
below\footnote{The present model cannot produce 
the correct 
%values for the
Weinberg angle, so that we take the four dimensional gauge coupling to
be a free parameter, and it is assumed to be of order of $O(1)$.}. 
%%%%%%%%%%%%%
$$
\begin{array}{|c|cccc|ccccccc|}
\hline
 & z_{adj}^{(+)} & z_{fd}^{(+)} & z_{adj}^{(+)s} & z_{fd}^{(+)s} &  
   z_{adj}^{(-)} & z_{fd}^{(-)}   & z_{adj}^{(-)s} & z_{fd}^{(-)s}& 
   {1\over{g_4R}} & a_0 & m_H/g_4^2 
\\ \hline\hline
(1)& 0 & 0 & \mbox{-} & \mbox{-} & 0 & 0 & 0 & \mbox{-} & 6.3
& 0.039 & 134.0 
\\ \hline
%
%(2)& 0.1 & 0.1 & \mbox{-} & \mbox{-} & 0 & 0 & 0 & \mbox{-} & 9.1
%& 0.027 & 136.5
%\\ \hline
%
(2)& 0.1 & 0.2 & \mbox{-} & \mbox{-} & 0 & 0 & 0 & \mbox{-} & 12.4
& 0.020 & 139.4
\\ \hline
(3)& 0 & 0 & \mbox{-} & \mbox{-} & 0 & 0 & 0.5 & \mbox{-} & 7.7
& 0.032 & 141.0
\\ \hline
%
%(3)& 0.1 & 0 & \mbox{-} & \mbox{-} & 0 & 0 & 0 & \mbox{-} & 8.4
%& 0.029 & 136.1
%\\ \hline
%(4)& 0 & 0.1 & \mbox{-} & \mbox{-} & 0 & 0 & 0 & \mbox{-} & 6.5
%& 0.038 & 134.8
%\\ \hline
%
%(5)& 0.1 & 0.1 & \mbox{-} & \mbox{-} & 0 & 0 & 0.4 & \mbox{-} & 12.3
%& 0.020 & 140.8
%\\ \hline
%
(4)& 0.1 & 0.2 & \mbox{-} & \mbox{-} & 0 & 0 & 0.2 & \mbox{-} & 14.0
& 0.018 & 140.4
\\ \hline
(5)& 0.1 & 0.2 & \mbox{-} & \mbox{-} & 0.1 & 0 & 0 & \mbox{-} & 10.4
& 0.024 & 137.2
\\ \hline
(6)& 0.1 & 0.2 & \mbox{-} & \mbox{-} & 0 & 0.2 & 0 & \mbox{-} & 11.0
& 0.022 & 137.9
\\ \hline
(7)& 0.2 & 0.2 & \mbox{-} & \mbox{-} & 0.3 & 0.3 & 0 & \mbox{-} & 24.0
& 0.010 & 109.1
\\ \hline
%%%%
%(7)& 0.1 & 0.2 & \mbox{-} & \mbox{-} & 0 & 0 & 0 & \mbox{-} & 12.4
%& 0.020 & 139.4
%\\ \hline
%(8)& 0.2 & 0.2 & \mbox{-} & \mbox{-} & 0.4 & 0 & 0 & \mbox{-} & 6.7
%& 0.037 & 116.9
%\\ \hline
%(9)& 0.5 & 0 & \mbox{-} & \mbox{-} & 0.9 & 0 & 0.7 & \mbox{-} & 8.0
%& 0.031 & 75.6
%\\ \hline
%
\end{array}
$$
\par
%%%%%%%%%
In the table (and the subsequence tables), the Higgs mass is 
measured in GeV unit and the radius of
the $S^1$ is in TeV unit. The case (1) corresponds to the massless bulk
fields.
\par
%%%%
%The case (2) is a typical example that the field with
%$\delta=1$ enhances the Higgs mass through the large logarithmic 
%factor due to the small VEV $a_0$, which is achieved by the
%cancelation of the coefficient $B$ thanks to the field with
%$\delta=1$. 
%%%%%%
The equation (\ref{shiki17}) implies that the bulk 
mass $z_{adj(fd)}^{(+)}$ tends not to make the Higgs mass heavy.
In order to have the heavy Higgs mass, the VEV $a_0$ for the Higgs
field has to be smaller. 
The case (2) just shows that the Higgs mass is enhanced 
by the smaller VEV for the moderate size of the introduced 
bulk mass $z_{adj(fd)}^{(+)}$.
\par
%%%%%%%
The cases (3) $-$ (6) are examples showing 
the effect of the bulk mass for the field with $\delta=1$ on the Higgs mass. 
From Eq. (\ref{shiki17}), the field with $\delta=1$ does not affect 
the Higgs mass directly, but it contributes to increase or decrease 
the Higgs mass, depending on how small the 
VEV $a_0$ for the Higgs field is induced due to the bulk mass $z^{(-)}$. 
The cases (3) and (4) show that the bulk 
mass $z_{adj}^{(-)s}$ makes the VEV small. 
As a result, the Higgs mass is enhanced compared with the 
corresponding cases with $z_{adj}^{(-)s}=0$. 
On the other hand, the cases (5) and (6) show that the bulk 
mass $z_{adj(fd)}^{(-)}$ decreases the Higgs mass 
because it does not induce the smaller VEV $a_0$ compared 
with the correspondeing cases with $z_{adj(fd)}^{(-)}=0$. 
\par
%%%%%%%%
%The bulk mass does not direcly affect the 
%Higgs mass, but thanks to the
%small VEV $a_0$ by the bulk mass for the field 
%with $\delta=1$, the Higgs mass is lifted. This 
%is clear if we compare (6) ((2)) with the case (5) ((1)).   
%\par
%%%%%%%%
%The typical effects of the bulk mass is manifest
%in the case (7). Thanks to the bulk mass, the VEV $a_0$ is 
%tuned to be small values. The bulk mass, however, suppresses
%the large logarithmic contributions by the reason stated 
%before, so that the Higgs mass cannot become heavy, expected 
%from the obatined small values for $a_0$. This is also clear if 
%we compare (1) with (5).
%\par
%%%%%
%The case (7)  the large
%suppression by the bulk mass. 
Although the very small VEV $a_0$ is 
realized in (7) by adjusting the bulk mass parameres, 
the effects of $\delta=0$ fermions with the large bulk masses 
dominate in the argument of the logarithm. 
%so that there is no
%ingredient to enhance the size of the Higgs mass.
That is why the Higgs mass is light in this case.
\par
%%%%%%%%%%%    
%Here, we note that the bulk mass also has a role to make 
%the bulk field heavy, decouple from low-energy 
%physics, in addition to the role stated in the 
%introduction. In this respect, the field 
%satisfying the antiperiodic boundary condition has a 
%lowest mass of order of $O(1/R)$, so that it is decoupled 
%from low-energies physics if $O(1/R)\sim$ TeV, which is 
%realized by the small VEV $a_0$. Therefore, for the matter 
%contents (B) and (C), we do not consider the bulk mass 
%for the antiperiodic field. 
\par
%%%%%%
Let us next present the results for the 
case (B), which is summarized
in the table,
$$
\begin{array}{|c|cccc|ccccccc|}
\hline
 & z_{adj}^{(+)} & z_{fd}^{(+)} & z_{adj}^{(+)s} & z_{fd}^{(+)s} &  
   z_{adj}^{(-)} & z_{fd}^{(-)}   & z_{adj}^{(-)s} & z_{fd}^{(-)s}& 
   {1\over{g_4R}} & a_0 & m_H/g_4^2 
\\ \hline\hline
(1)& 0 & 0 & \mbox{-} & 0 & 0 & 0 & 0 & \mbox{-} & 4.0
& 0.062 & 117.4 
\\ \hline
(2)& 0.2 & 0.2 & \mbox{-} & 0.2 & 0 & 0 & 0 & \mbox{-} &10.4
& 0.024 & 118.6
\\ \hline
(3)& 0.2 & 0.1 & \mbox{-} & 0.2 & 0 & 0 & 0 & \mbox{-} & 7.7
& 0.032 & 120.0
\\ \hline
%
%(4)& 0.2 & 0 & \mbox{-} & 0.2 & 0 & 0 & 0 & \mbox{-} & 7.2
%& 0.034 & 120.0
%\\ \hline
%
(4)& 0.2 & 0.25 & \mbox{-} & 0.2 & 0 & 0 & 0 & \mbox{-} & 15.5
& 0.016 & 115.0
\\ \hline
(5)& 0.2 & 0.3 & \mbox{-} & 0.5 & 0 & 0 & 0 & \mbox{-} & 7.5
& 0.033 & 118.2
\\ \hline
\end{array}
$$
%Again, the Higgs mass is measured in GeV unit and the radius of
%the $S^1$ is in TeV unit. The case (1) corresponds to the massless bulk
%fields. 
In these cases, the Higgs mass slightly small compared 
with the previous cases though the VEV $a_0$ is almost the same values. 
%as the former cases. 
This is clear from the fact that the all the fields with $\delta=0$ 
have the bulk mass, so that the logarithmic factor, which could 
make the Higgs mass
heavy, cannot be large due to the increased argument in the logarithm.
\par
%%%%%%%%%%%%%%%%%%%%%%%%%%%
Finally let us present the numerical results for the case (C),
which is given by
$$
\begin{array}{|c|cccc|ccccccc|}
\hline
 & z_{adj}^{(+)} & z_{fd}^{(+)} & z_{adj}^{(+)s} & z_{fd}^{(+)s} &  
   z_{adj}^{(-)} & z_{fd}^{(-)}   & z_{adj}^{(-)s} & z_{fd}^{(-)s}& 
   {1\over{g_4R}} & a_0 & m_H/g_4^2 
\\ \hline\hline
(1)& 0 & 0 & \mbox{-} & \mbox{-} & 0 & 0 & \mbox{-} & 0 & 6.1
& 0.040 & 130.2 
\\ \hline
(2)& 0.1 & 0 & \mbox{-} & \mbox{-} & 0 & 0 & \mbox{-} & 0 & 8.2
& 0.030 & 132.6
\\ \hline
(3)& 0.1 & 0.1 & \mbox{-} & \mbox{-} & 0 & 0 & \mbox{-} & 0 & 8.9
& 0.028 & 133.2
\\ \hline
(4)& 0.1 & 0.2 & \mbox{-} & \mbox{-} & 0 & 0 & \mbox{-} & 0 & 12.2
& 0.020 & 136.3
\\ \hline
(5)& 0 & 0.4 & \mbox{-} & \mbox{-} & 0 & 0 & \mbox{-} & 0 & 19.2
& 0.013 & 161.0
\\ \hline
(6)& 0.09 & 0.3 & \mbox{-} & \mbox{-} & 0 & 0 & \mbox{-} & 0 & 28.4
& 0.0087 & 139.7
\\ \hline
(7)& 0.08 & 0.3 & \mbox{-} & \mbox{-} & 0 & 0 & \mbox{-} & 0 & 17.3
& 0.014 & 143.2
\\ \hline
\end{array}
$$
%%%%%%%%%%
%All the bulk mass is massless in the case (1), and the Higgs 
%mass (the radius of the $S^1$) is in the GeV (TeV) unit. 
The case (6) (also the cases (5), (7) and (7) for the case (A)) is 
an example that the bulk mass is finely tuned 
for the coefficient of the quadratic terms for $a^2$ 
to yield the very small values of the VEV $a_0$. 
In this case, one needs to take care whether or not the very small
values of $a_0$ is reliable result within the framework of 
perturbation theory. 
%In fact, roughly speaking, 
Since the number of order $O(10^{-3})$ is smaller or 
at least comparable to the loop factor, one needs to check 
the stability of obtained result against higher loop corrections.
\par
%%%%%%
In the case (5), if we take the color factor $3$ into account 
\footnote{In this case, we start with 
the $SU(3)_c\times SU(3)$ group and set the orbifolding 
boundary condition for the $SU(3)_c$ is trivial.}, the 
Higgs mass $m_H\simeq 120$ GeV is consistent with the
experimental lower bound even if we take 
the four dimensional gauge coupling to be the $SU(2)$ gauge 
coupling. The rough estimation tells us that in order 
for $\sqrt{3}g_{SU(2)}^2 Q^2$, which is equivalent to
the square of the effective coupling $g_{eff}^2$, to 
be $O(1)$, it is enough to introduce the matter 
with $Q=j=3/2$, that is, ${\bf 10}$ of $SU(3)$.
%%%%%%%%%%%%%%%%%%%%%%%%%%%%%%%%%%%%%
\section{Conclusions and discussions}
%%%%%%%%%%%%%%%%%%%%%%%%%%%%%%%%%%%%%
We have considered the five dimensional $SU(3)$ gauge-Higgs unification 
theory defined on $M^4\times S^1/Z_2$. 
Its matter content includes a pair of the fields satisfying 
the periodic and the antiperiodic boundary conditions. 
Then, the parity even mass terms composed of these fields
are introduced. 
%by the fields. 
We have studied the effect 
of the bulk mass in the scenario of the
gauge-Higgs unification, in which 
the bulk mass is important when we discuss 
the realistic fermion mass spectrum. 
\par
%%%%%%%
In particular, we have investigated 
the effect of the bulk mass on the vacuum expectation 
values for the Higgs field and the size of the Higgs mass. 
The analyses have been performed using the transparent and useful 
expressions (\ref{newshiki1}) and (\ref{shiki17}), 
%obtained by using the formulae (\ref{shiki15}) and (\ref{shiki16}), 
which helps us to understand the effect of the bulk mass clearer. 
\par
%%%%%%%%
We have found that the VEV for the Higss field $a_0$ can be small, 
which is necessary for the scenario of the gauge-Higgs unification, 
due to the bulk mass parameters for a fixed set of matter content.
The effect of the bulk mass on the VEV is different, 
depending on the periodicity of the field. 
We have found in our numerical analyses that the bulk 
mass $z_{adj}^{(-)s} (z_{adj(fd)}^{(-)})$ is likely 
to make the VEV smaller (larger).  
%and observed 
%that the Higgs mass is enhanced by the bulk mass $z_{adj}^{(-)s}$.  
\par
%%
%The field 
%with $\delta=1$, the antiperiodic field, plays an 
%important role in doing it. 
%
%%%%%%%
The size of the Higgs mass is also affected by the
bulk mass parameters. The dominant contribution to the
Higgs mass comes from only the field with $\delta=0$.
The bulk mass for the field with $\delta=0$, however, tends to 
make the Higgs mass less heavy because it increases the 
argument of the logarithm in Eq.(\ref{shiki17}). One needs 
smaller values for the VEV in order to obtain 
the heavy Higgs mass for the moderate size of the bulk mass.
On the other hand, the direct effect of the bulk mass 
for the field with $\delta=1$ on the Higgs mass is 
tiny, but it has a role to make the VEV large or small. 
%
%and accordingly, 
%
We have especially observed that 
the Higgs mass is enhanced (decreased) by the bulk 
mass $z_{adj}^{(-)s} (z_{adj(fd)}^{(-)})$. 
%%%%%
%This implies that the Higgs mass is affected 
%through the size of the VEV, which is 
%affected by the bulk mass for the field with $\delta=1$.
%%%%
\par
%%%%%%%
%We have known that the bulk mass is important 
%when we discuss the realistic fermion spectrum. 
%We have clarified in this paper that the bulk mass
%also plays crucial role to have the correct electroweak 
%symmetry breaking (small VEV $a_0$) and size of the Higgs mass. 
\par
%%%%%%%
For some cases that we have numerically studied, the VEV $a_0$ becomes 
the order of $O(10^{-3})$. In this case, one needs to take care 
the stability of obtained result against higher order loop corrections 
to the effective potential since the order is 
almost the same as the loop factor. 
In this respect, as in the case with 
the matter in the higher dimensional representation 
under the gauge group for which the radiative 
corrections is sizable \cite{csaki}, we need to investigate the 
two loop effects to the effective potential when we have such 
the small VEV.
\par
%%%%%%%%% 
%So far, we have understood the effects of the bulk mass parameter on the
%gauge-Higgs unification. 
We have clarified in this paper that the bulk mass plays crucial roles 
to have the correct electroweak symmetry breaking (small VEV $a_0$) 
and the size of the Higgs mass. 
In connection with the electroweak phase transition 
at finite temperature through the dynamics of the Wilson line
phases \cite{panico, mt2}, it is very interesting to study 
the effect of the bulk mass on the order of the phase transition. 
In particular, it is important to understand how the bulk 
mass affects the first order phase transition, which is necessary 
for the scenario of the electroweak baryogenesis. 
This will be reported soon \cite{mt}.   
\vspace{1cm}
%\newpage
%%%%%%%%%%%%%%%%%%%
\begin{center}
{\bf Acknowledgements}
\end{center}  
N.M. is supported by I.N.F.N., sezione di Roma. K.T. is supported by
the $21$st Century COE Program at Osaka University and would like to 
thank Professor Y. Hosotani for fruitful discussions. 
He would also like to thank Dr. T. Yamashita 
for useful comments on the gauge-Higgs unification. K.T. would
also like to thank the Yukawa Institute for Theoretical 
Physics for warm hospitality, where a part of this 
work has done.
%%%%%%%%%%%%%%%%%%%%%%%%%%%%%
\vspace*{1cm}
\begin{center}
{\bf Appendix}
\end{center}  
In this appendix, 
we briefly explain how the expressions (\ref{shiki15}) 
and (\ref{shiki16}) are derived. 
The function $f(Qa, z, \delta)$ is also written as
\begin{eqnarray}
f(Qa, z, \delta)&=&{\rm Re~Li}_5
\biggl(\e^{-z+2\pi i (Qa-{\delta\over 2})}\biggr)
+z~{\rm Re~Li}_4
\biggl(\e^{-z+2\pi i (Qa-{\delta\over 2})}\biggr)
\nonumber\\
&+&{z^2\over 3}{\rm Re~Li}_3 
\biggl(\e^{-z+2\pi i (Qa-{\delta\over 2})}\biggr),
\end{eqnarray}
where the polylogarithm function is defined by 
\begin{equation}
{\rm Li}_D(\e^{-z}) = \sum_{n=1}^{\infty}{\e^{-nz}\over n^D}.
\end{equation}
The asymptotic form for $z \ll 1$ is given by \cite{dj}
\begin{eqnarray}
{\rm Li}_D(\e^{-z})
&\simeq& \sum_{n=0 (n\neq D-1)}^{\infty}(-1)^n {z^n\over n!}\zeta(D-n)
\nonumber\\
&+&(-1)^D {z^{D-1}\over{(D-1)!}}\biggl({\rm ln}z -
\Bigl(1+\half+{1\over 3}+\cdots +{1\over{D-1}}\Bigr)\biggr),
\\ 
{\rm Li}_D(-\e^{-z})&\simeq& 
\sum_{n=0}^{D-2}(-)^{n+1}{z^n\over {n!}}\eta(D-n) + 
(-)^D {{z^{D-1}}\over{(D-1)!}}{\rm ln}2
+(-)^{D+1}\half {{z^D}\over {D!}}\nonumber\\
&+&(-)^D\sum_{m=1}^{\infty}{{z^{D+2m-1}}\over {(D+2m-1)!}}~\eta(1-2m),
\end{eqnarray}
where $\eta(s)\equiv (1-2^{1-s})\zeta(s)$. In order to obtain the 
expressions (\ref{shiki15}) and (\ref{shiki16}),
we first expand the $\cos(na)$ with respect to $a$ since $a$ is small. 
%in terms of polynomial.
%And then, 
In addition to the formulae given above, 
using the useful expressions
\begin{eqnarray}
\sum_{n=1}^{\infty}{1\over n}\e^{-nz}&=&
{\rm ln}\left(1-\e^{-z}\right)^{-1}\simeq 
-{\rm ln}(z)+{z\over 2}-{z^2\over 24}
+{z^4\over 2880}-{z^6\over 181440}+O(z^8),
\label{shikiap1}\\
\sum_{n=1}^{\infty}{(-1)^n\over n}\e^{-nz}
&=&{\rm ln}\left(1+\e^{-z}\right)^{-1}
\simeq -{\rm ln}2+{z\over 2}-{z^2\over 8}
+{z^4\over 192}-{z^6\over 2880}+O(z^8)
\label{shikiap2}
\end{eqnarray}
and %the expressions obtained by taking 
the derivatives of (\ref{shikiap1}) and (\ref{shikiap2}) 
with respect to $z$ lead to (\ref{shiki15}) and (\ref{shiki16}) in the text.  
%%%%%%%%%%%%%%%%%%%
%%%%%%%%%%%%% BIBLIOGRAPHY %%%%%%%%%%%%%%%%%%%%

%%%%%%%%%%%%%%%%%%%%%%%%%%%%
\end{document}